\documentclass[conference]{IEEEtran}
\usepackage{algorithmic}
\usepackage{array}
\usepackage{tikz}
\usetikzlibrary{positioning,calc,shapes.misc}
\usetikzlibrary{arrows.meta}
\usepackage{amsmath,amssymb,braket}
\usepackage{booktabs}
\usepackage{subcaption}
\usepackage{enumitem}
\usepackage{xcolor}
\usepackage[T1]{fontenc}

\IEEEoverridecommandlockouts

\ifCLASSOPTIONcompsoc
 \usepackage[caption=false,font=footnotesize,labelfont=sf,textfont=sf]{subfig}
\else
 \usepackage[caption=false,font=footnotesize]{subfig}
 \usepackage{stmaryrd}
 \usepackage{subcaption}

\fi

\newcommand{\quantparam}[1]{[\![#1]\!]}

\newcommand{\ma}[1]{\textcolor{black}{#1}}

\begin{document}

\title{Resource-Adaptive Teleportation Under Imperfect Entanglement:
A Code-Puncturing Framework}

\author{
  \IEEEauthorblockN{
    Mahmoud Saad Abouamer, 
    Jaron Skovsted Gundersen, 
    Søren Pilegaard Rasmussen, 
    Petar Popovski}
  \IEEEauthorblockA{
    Department of Electronic Systems, Aalborg University,
    9220 Aalborg, Denmark\\
    Email: \{mahmoudabo, jaron, spr, petarp\}@es.aau.dk}
  \thanks{This work was supported, in part, by the Danish National Research Foundation (DNRF),
      through the Center CLASSIQUE, grant nr.\ 187.}
}

\maketitle
\begin{abstract}
Quantum teleportation is a foundational protocol for sending quantum information through entanglement distribution and classical communication. Assuming ideal classical communication, the reliability of quantum teleportation is limited by the fidelity of the shared EPR pairs. This reliability can be improved through two mechanisms: entanglement purification and quantum error correction (QEC). Using both techniques in concert requires flexible QEC rates, since purification alters the structure of errors induced by imperfect-EPR teleportation, and fixed-rate codes cannot be uniformly effective across purification regimes or reliability targets. In this work, we supplement purification with punctured QEC codes, providing a family of code variants that can be adapted to error-channel characteristics and reliability targets. Punctured codes improve teleportation reliability across a broader range of purification regimes, enabling target reliability to be met without hardware-level code switching. This is corroborated by numerical results, showing that different punctured codes achieve the lowest logical error probability in different operating regimes, and that selecting among them reduces logical error relative to fixed-rate encoded teleportation. This reduction relaxes the requirement on the initial EPR fidelity or purification needed to achieve a target reliability. Overall, puncturing enables adaptation to varying entanglement conditions and reliability requirements while reusing a single stabilizer structure.

\end{abstract}

\IEEEpeerreviewmaketitle

\section{Introduction}
\label{introduction section}

\begin{figure}[t]
\centering

\begin{minipage}{0.3\textwidth}
    \centering
    \includegraphics[width=\linewidth]{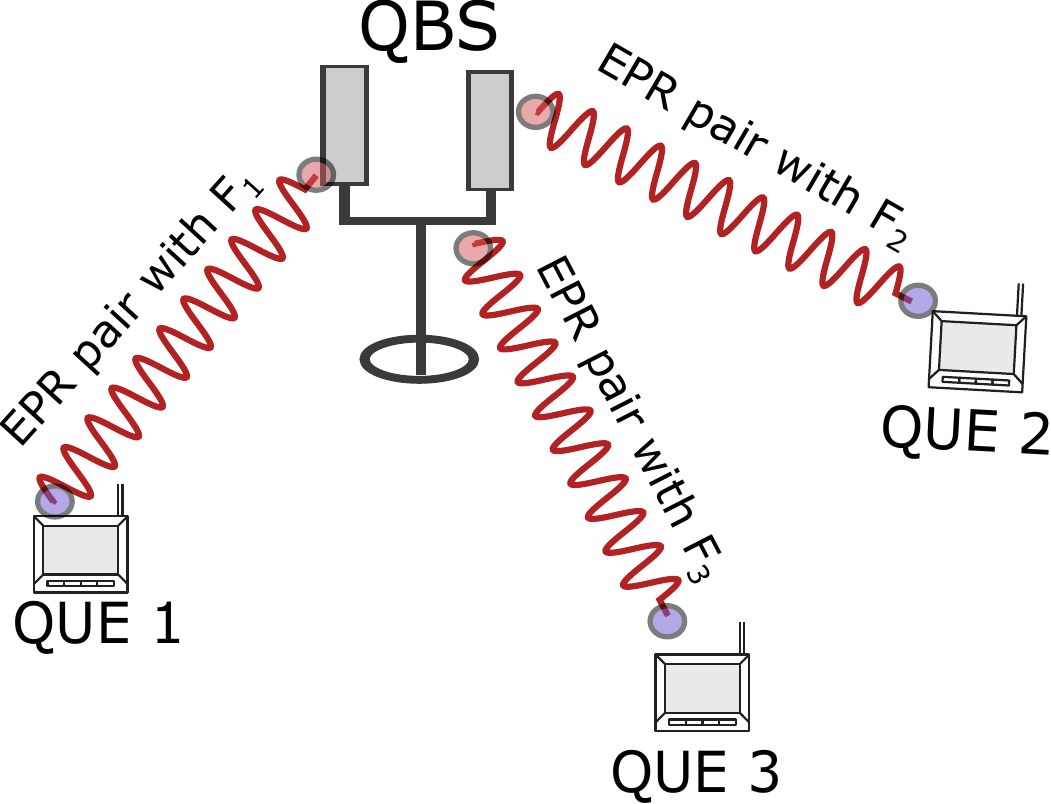}
\end{minipage}\hfill

\begin{minipage}{0.49\textwidth}
    \centering
\includegraphics[width=\linewidth]{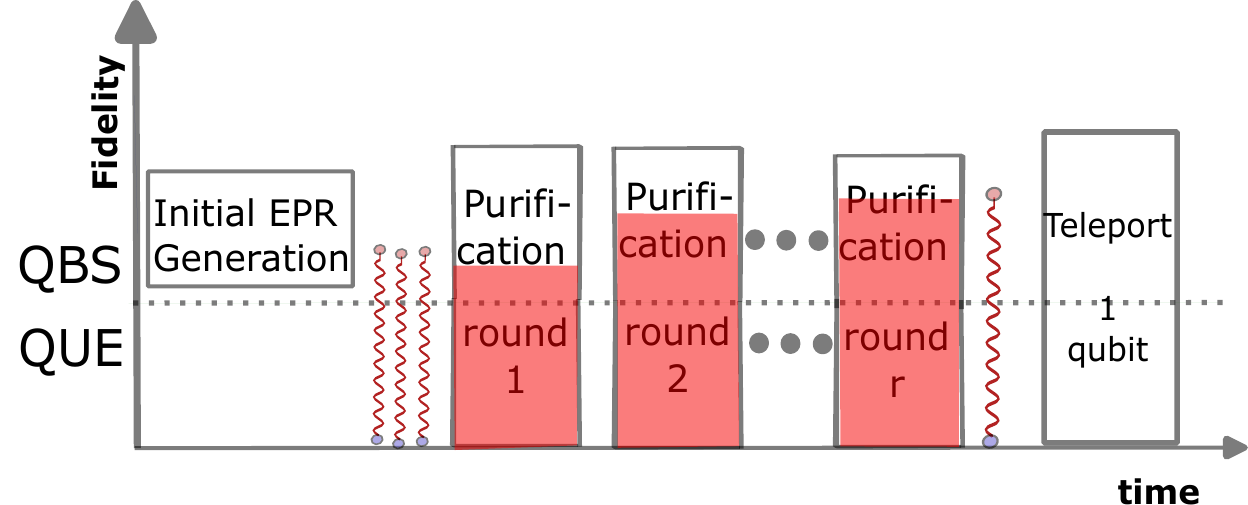}
\end{minipage}

\caption{Heterogeneous reliability in quantum teleportation: 
(top) a 1Q network with user-dependent EPR fidelities; 
(bottom) a schematic timing diagram showing the competition between purification and application time prior to teleportation.}
\label{1Q_like_figure}
\end{figure}

Quantum teleportation is a key primitive for distributed quantum computing and secure networking. Its performance, however, is fundamentally limited by the fidelity of the shared Einstein–Podolsky–Rosen (EPR) pairs. In emerging multi-user architectures, such as the 1Q framework~\cite{1Qpaper} illustrated in Fig.~\ref{1Q_like_figure}, a quantum base station (QBS) distributes entanglement to multiple quantum user equipments (QUEs) over heterogeneous links. Differences in physical media, path length, and routing strategies lead to substantial variation in EPR fidelity across users~\cite{kumar2024routing}. Consequently, teleportation reliability is inherently non-uniform and varies across both users and time. This motivates adaptive mechanisms that account for heterogeneous entanglement quality and application-specific reliability requirements.

Entanglement purification can mitigate EPR fidelity degradation by distilling multiple noisy pairs into fewer, higher-quality ones using local operations and classical communication. However, purification incurs additional latency due to repeated purification rounds, and classical coordination, during which stored entanglement continues to decohere~\cite{pur_with_latency_arxiv,pur_with_latency_pre_conf}. Consequently, additional purification can yield diminishing or even negative returns, and strict latency budgets often make deep purification infeasible. In time-sensitive or resource-limited settings, purification alone may therefore be insufficient to achieve the required reliability targets, motivating the use of QEC as a complementary mechanism.

While unencoded teleportation cannot outperform the fidelity of the shared EPR pair, encoding introduces structured redundancy that enables recovery at the decoder, allowing target reliability levels to be met even when input fidelity is modest. Encoded teleportation therefore combines purification with QEC to suppress logical error beyond what is achievable using unencoded teleportation, as investigated in~\cite{Lorenzo2024}. However, the effectiveness of QEC depends not only on the target reliability but also on the structure of the error channel induced by teleportation with imperfect EPR pairs. In particular, purification alters not only the overall fidelity but also introduces asymmetries between different error components. The resulting changes typically increase with purification and can differ across users, leading to heterogeneous teleportation channels that vary in both quality and error structure.

These effects highlight a key limitation of fixed-rate QEC for teleportation: no single code is uniformly effective across purification regimes, error-channel asymmetries, or reliability targets. However, switching between unrelated stabilizer codes is undesirable in practical deployments, as it requires changes to stabilizer measurements, encoder/decoder implementations, and often hardware-level support at the QBS or QUE. \ma{This motivates adaptive coding approaches that adjust error protection to channel conditions while reusing a common underlying stabilizer structure. This can be achieved through a quantum analogue of classical rate-compatible punctured coding, where a single low-rate “mother” code is punctured to obtain a family of codes that reuse the same encoder and decoder~\cite{Hagenauer_punc}.}

\ma{In this work, we leverage puncturing of stabilizer codes to enable resource-adaptive encoded teleportation. Building on prior analyses of punctured code validity~\cite{Puncturing}, we puncture a single Calderbank–Shor–Steane (CSS) code to generate a family of related codes with varying rates and asymmetric $X/Z$ error-protection profiles, all sharing a common stabilizer structure. This allows teleportation performance to be tuned to different EPR fidelities, error-channel characteristics, and reliability targets while retaining a single stabilizer framework and without modifying encoder or decoder hardware. \textcolor{black}{In particular, the shortest punctured code that satisfies the reliability target can be selected to minimize resource usage. Numerical results  corroborate the efficacy of the proposed framework,} showing that punctured codes improve teleportation reliability and reduce logical error compared to fixed-rate encoding, achieving target reliability with lower initial EPR fidelity or fewer purification rounds.}

\begin{figure*}[t]
\centering
\begin{tikzpicture}[
    node distance=1cm,
    every node/.style={
        draw,
        fill=gray!20,
        rounded corners,
        text width=3.5cm,   
        minimum height=1.6cm,
        align=center,
        font=\small,
    },
    arrow/.style={->, very thick}
]

\node (enc) {
    \textbf{Quantum\\Encoder}\\[2pt]
    (Base CSS code)\\[4pt]
    $\quantparam{n_{\text{base}},\, k,\,
      d^{\text{base}}_X / d^{\text{base}}_Z }$
};

\node (punc) [right=of enc] {
    \textbf{Puncturing}\\[4pt]
    $\quantparam{ n_{\text{punc}},\, k,\,
      d^{\text{punc}}_X / d^{\text{punc}}_Z }$
};

\node (pauli) [right=of punc] {
    \textbf{ Imperfect Teleportation Pauli Channel ($n_{\text{punc}}$ uses})\\[4pt]
    $p_X^{(r)} = C_r,\quad
     p_Y^{(r)} = B_r,\quad
     p_Z^{(r)} = D_r$
};

\node (dec) [right=of pauli] {
    \textbf{Decoding and\\error correction}\\[2pt]
    logical error $P_L^{(r)}$
};

\draw[arrow] (enc) -- (punc);
\draw[arrow] (punc) -- (pauli);
\draw[arrow] (pauli) -- (dec);

\end{tikzpicture}

\caption{Overall coding-puncturing-purification framework. Punctured CSS codes set correction radii $(t_X, t_Z)$; together with the Pauli channel described in Section.~\ref{subsec:Imperfect Teleportation-induced Pauli Channel} and the resulting logical \textcolor{black}{error probability} $P_L^{(r)}$.}
\label{fig:pipeline}
\end{figure*}
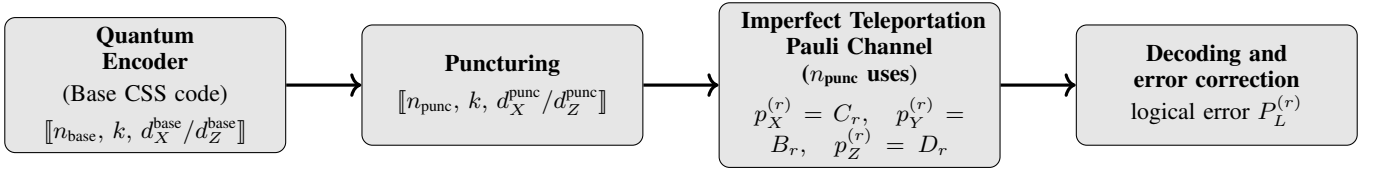

\section{Imperfect Teleportation-Induced Channel and Logical Error Probability}
\label{modelling section}

In this section, teleportation using imperfect EPR pairs is captured by a Pauli channel with parameters that depend on the initial EPR fidelity and the applied purification rounds. By combining this with the effects of quantum error correction, this serves as the foundation for evaluating logical error probabilities, which constitute our reliability metric, under stabilizer and CSS codes.
\subsection{\textcolor{black}{Imperfect Teleportation-induced Pauli Channel}}
\label{subsec:Imperfect Teleportation-induced Pauli Channel}
In the ideal case, teleportation through a perfect Bell pair $\ket{\Phi^+}$ transmits the quantum state without disturbance. When entanglement is imperfect, teleportation introduces Pauli byproduct errors determined by the Bell-state decomposition of the shared resource~\cite{Lorenzo2024}: 
\begin{align}
\rho &= A\ket{\Phi^+}\bra{\Phi^+} 
     + B\ket{\Psi^-}\bra{\Psi^-} \nonumber\\
     &\quad + C\ket{\Psi^+}\bra{\Psi^+} 
     + D\ket{\Phi^-}\bra{\Phi^-}.
\end{align} 
In the teleportation circuit, the Bell components $\{\ket{\Phi^+}, \ket{\Psi^-}, \ket{\Psi^+}, \ket{\Phi^-}\}$ correspond to Pauli operators $\{I, Y, X, Z\}$. Thus, teleporting through the state $\rho$ is equivalent to a Pauli channel with \textcolor{black}{error probabilities}:
\begin{equation}
p_X = C, \qquad p_Y = B, \qquad p_Z = D.
\label{eq:teleportation_pauli_channel}
\end{equation}

We now turn to how purification modifies the Pauli channel, increasing fidelity while often introducing asymmetry between error types. Following~\cite{Dur2007EntanglementPurification, Lorenzo2024}, initial EPR pairs are taken to be Werner states with fidelity $F_0$, resulting in a symmetric Pauli channel with $p_X^{(0)} = p_Y^{(0)} = p_Z^{(0)} = \frac{1 - F_0}{3}$. \ma{Here, $F_0$ captures the quality of the shared EPR pairs, and its impact on logical error probability is evaluated by sweeping $F_0$ in Section~\ref{sec:Numerical Evaluations}.}

Purification protocols can increase fidelity by consuming multiple noisy EPR pairs and applying local operations with classical communication (LOCC). As a concrete scheme, we consider the DEJMPS protocol~\cite{Deutsch1996DEJMPS}. 
After $r$ purification rounds under DEJMPS, the Bell coefficients evolve to $(A_r, B_r, C_r, D_r)$ according to:
\begin{align}
A_{r+1} &= (A_r^2 + B_r^2)/N_r, \qquad 
B_{r+1} = 2C_r D_r / N_r, \nonumber \\
C_{r+1} &= (C_r^2 + D_r^2)/N_r, \qquad 
D_{r+1} = 2A_r B_r / N_r,
\end{align}
with normalization $N_r = (A_r + B_r)^2 + (C_r + D_r)^2$~\cite{Deutsch1996DEJMPS}.
This induces the following updated Pauli error probabilities:
\begin{align}
   p_X^{(r)} = C_r, \qquad
p_Y^{(r)} = B_r, \qquad
p_Z^{(r)} = D_r. 
\end{align}

These \textcolor{black}{expressions} allow us to compute the Pauli error \textcolor{black}{probabilities}  at each purification round and quantify how asymmetry develops over time. As shown in~\cite{Lorenzo2024}, purification can introduce significant imbalance between $X$ and $Z$ errors, especially at moderate initial fidelity. The extent of this asymmetry depends on the number of rounds, the protocol used, and the input state.

\ma{We note that, although more sophisticated entanglement purification protocols have been proposed since DEJMPS, the specific choice of purification scheme plays no fundamental role here. Purification enters solely through its effect on the induced Pauli error probabilities, which fully characterize the teleportation channel. Consequently, other purification protocols can be similarly incorporated, and DEJMPS is adopted here as a representative example.}

\subsection{Logical Error Probability under Quantum Error Correction}
With the Pauli \textcolor{black}{error probabilities} $(p_X^{(r)}, p_Y^{(r)}, p_Z^{(r)})$ determined from the shared EPR pair, we evaluate their effect on teleportation reliability when using quantum error correction. Encoded teleportation applies a stabilizer code to protect the transmitted qubit. The code's effectiveness depends on its interaction with the induced error channel.

\textcolor{black}{CSS codes are well-suited for Pauli noise because they decouple $X$ and $Z$ correction. Following~\cite{CSS_Error_Prob}, each branch is treated as a binary symmetric channel (BSC) with
\begin{equation}
q_X^{(r)} = p_X^{(r)} + p_Y^{(r)}, \qquad
q_Z^{(r)} = p_Z^{(r)} + p_Y^{(r)},
\end{equation}
corresponding to bit-flip and phase-flip error probabilities.}

For a CSS code of length $n$ with distances $d_X$ and $d_Z$, let $t_X = \lfloor (d_X - 1)/2 \rfloor$ and $t_Z = \lfloor (d_Z - 1)/2 \rfloor$ be the correction radii. The decoder succeeds if the number of errors on each branch is below its threshold:
\begin{align}
P_{\mathrm{succ},X}^{(r)} &= \sum_{i = 0}^{t_X} \binom{n}{i} (1 - q_X^{(r)})^{n - i} (q_X^{(r)})^i, \\
P_{\mathrm{succ},Z}^{(r)} &= \sum_{i = 0}^{t_Z} \binom{n}{i} (1 - q_Z^{(r)})^{n - i} (q_Z^{(r)})^i.
\end{align}
The total logical error probability is then
\begin{equation}
P_L^{(r)} = 1 - P_{\mathrm{succ},X}^{(r)} P_{\mathrm{succ},Z}^{(r)}.
\label{eq:PL_css_corrected}
\end{equation}

The coding-puncturing-purification framework is illustrated in Fig.~\ref{fig:pipeline}.   In the next section, we introduce puncturing to derive a family of related codes from a single CSS code. These variants have different rates and distinct error protection profiles, while sharing a common stabilizer structure. This construction allows encoded teleportation to adapt to different channel conditions without changing the stabilizer structure.

\section{Puncturing }
\label{puncturing section}

\subsection{CSS codes and Puncturing}
It is possible to define a quantum code from two classical codes $C_1$ and $C_2$, where $C_2^\perp\subseteq C_1$. If $C_1$ is an $[n,k_1]$ code and $C_2$ is an $[n,k_2]$ code, the CSS code is an $\quantparam{n,k_1+k_2-n,d_X/d_Z}$ asymmetric code defined from the stabilizer matrix
\begin{align}
    \begin{bmatrix}
        H_1 & 0\\
        0 & H_2
    \end{bmatrix},
\end{align}
where $H_i$ is the parity check matrix for $C_i$. We will only focus on codes which have dimension $1$ implying $k_1+k_2=n+1$. The minimum distances $d_X$ and $d_Z$ are determined by
\begin{align}
    d_X&= \min w_H(C_1\setminus C_2^\perp), \nonumber\\
    d_Z&= \min w_H(C_2\setminus C_1^\perp),
\end{align}
representing that we can correct $t_X = \lfloor (d_X - 1)/2 \rfloor$ $X$ errors and $t_Z = \lfloor (d_Z - 1)/2 \rfloor$ $Z$ errors. A codeword $\ket{\psi}_L$ is an $n$-qubit state which satisfies that for a row $\mathbf{a}$ in $H_1$ it holds that $X^{a_1}\otimes X^{a_2}\otimes\cdots\otimes X^{a_n}\ket{\psi}_L=\ket{\psi}_L$. Similarly, for a row $\mathbf{b}$ in $H_2$ we have that $Z^{b_1}\otimes Z^{b_2}\otimes\cdots\otimes Z^{b_n}\ket{\psi}_L=\ket{\psi}_L$. These operators are called stabilizers of the code. We will denote by $S^X_i$ and $S^Z_i$ the stabilizers corresponding to the $i$'th row of $H_1$ and $H_2$, respectively. The idea behind decoding a CSS code is that if an $X$ error had occurred it will anti-commute with at least one of the $S^Z_i$ stabilizers. On the other hand, if a $Z$ error occurred it will anti-commute with at least one of the $S^X_i$ stabilizers. Thus, if we measure the eigenvalues of the stabilizers we can, by checking which stabilizers have $-1$ as an eigenvalue, determine and correct the error. If all stabilizers have $1$-eigenvalue, either no error occurred or too many errors occurred such that we cannot correct.

The process of puncturing a stabilizer code from the stabilizer matrix is described in \cite{Puncturing} and in \cite{Grassl_2023} it is also described how a puncturing of a CSS code can be obtained from puncturing and shortening the classical codes used in the CSS construction. Following the approach in \cite{Puncturing} we say that we puncture the code with respect to $(0|1)$ on qubit $i$ when we, by using row operations, make sure that all except at most one row of the stabilizer matrix will lie in the span of $(0|1)$ when restricted to the $i$'th and $(i+n)$'th columns. Afterwards, the row not in the span is removed and the columns $i$ and $i+n$ is also removed. Note that this is equivalent to puncturing $C_1$ and shortening $C_2$ on index $i$, which is the approach in \cite{Grassl_2023}. Similarly, we can puncture w.r.t. $(1|0)$. This corresponds to shortening $C_1$ and puncturing $C_2$. In any case, one of the minimum distances will remain unchanged, while the other may decrease by one.

\subsection{Considered Quantum Codes and Puncturing Strategy}
\label{subsec:codes}

To evaluate the impact of imperfect entanglement and purification on the 
teleportation-based transmission scheme, we consider two sets of codes:
a small benchmark code (the $\quantparam{7,1,3/3}$ Steane code) and a larger code
that supports asymmetric puncturing (a \(\quantparam{17,1,5/5}\) known as the $4.8.8$ color code).

The \(\quantparam{17,1,5/5}\) code is a CSS code with 
\setcounter{MaxMatrixCols}{20}
\begin{align*}
&H_1 = H_2 = \\
    &\begin{bmatrix}
        1&1&0&1&1&0&1&0&1&0&1&0&0&0&0&1&0\\
        0&1&1&0&0&0&1&1&0&0&1&1&0&0&1&1&0\\
        0&0&1&1&1&0&0&0&0&0&0&0&0&0&1&0&0\\
        0&0&0&1&0&0&0&0&0&0&0&0&0&1&1&1&0\\
        0&0&0&0&1&1&1&0&0&1&0&0&1&1&1&0&1\\
        0&0&0&0&0&1&0&1&0&0&0&1&1&0&0&0&0\\
        0&0&0&0&0&0&1&1&1&1&1&0&1&1&0&1&0\\
        0&0&0&0&0&0&0&1&0&1&0&1&0&0&0&0&1
    \end{bmatrix}
\end{align*}

To our knowledge, this is the smallest known CSS code with $k=1$ and $d_X=d_Z=5$. Due to these minimum distances we are able to correct two $X$ errors and $Z$ errors. We want to reduce $d_X$ to $3$ without reducing $d_Z$ by puncturing multiple times. This can be done by puncturing the code with respect to the element $(0|1)$ at different indices. This is also equivalent to puncturing $C_1$ (and hence shortening $C_1^\perp$) and shortening $C_2$ (and hence puncturing $C_2^\perp$). We found that we could puncture $4$ qubits obtaining an \(\quantparam{13,1,3/5}\) code by shortening the space spanned by the rows in $H_1$ $(C_1^\perp)$ and puncturing the space spanned by the rows in $H_2$ $(C_2^\perp)$ on the indices $1,2,9,11$. Performing these operations give us a CSS code with

\begin{align*}
H_1 &=
\begin{bmatrix}
1&1&1&0&0&0&0&0&0&0&1&0&0\\
0&1&0&0&0&0&0&0&0&1&1&1&0\\
0&0&1&1&1&0&1&0&1&1&1&0&1\\
0&0&0&1&0&1&0&1&1&0&0&0&0\\
0&0&0&0&0&1&1&1&0&0&0&0&1\\
\end{bmatrix}, \\
H_2 &=
\begin{bmatrix}
0&1&1&0&1&0&0&0&0&0&0&1&0\\
1&0&0&0&1&1&0&1&0&0&1&1&0\\
1&1&1&0&0&0&0&0&0&0&1&0&0\\
0&1&0&0&0&0&0&0&0&1&1&1&0\\
0&0&1&1&1&0&1&0&1&1&1&0&1\\
0&0&0&1&0&1&0&1&1&0&0&0&0\\
0&0&0&0&1&1&1&0&1&1&0&1&0\\
\end{bmatrix}.
\end{align*}

Note that this procedure has removed $S^X_1$, $S^X_2$, $S^X_7$, and $S^Z_8$ as stabilizers before removing the qubits. 

Afterwards, we puncture such that $d_Z$ is reduced to $3$ without affecting $d_X$. This can be done by puncturing with respect to the element $(1|0)$. Here we interchange puncturing and shortening on the classical codes, such that we now puncture the span of the rows in $H_1$ and shorten the span of the rows in $H_2$. We found that by performing these operations on the indices $4,6,7,8,9$, (or indices $6,8,10,12,13$ in the original code) we obtain a CSS code with
\begin{align*}
H_1 &=
\begin{bmatrix}
1&1&1&0&0&1&0&0\\
0&1&0&0&1&1&1&0\\
0&0&1&1&1&1&0&1\\
0&0&0&0&0&0&0&1
\end{bmatrix},\\[6pt]
H_2 &=
\begin{bmatrix}
0&1&1&1&0&0&1&0\\
0&1&0&0&1&1&1&0\\
1&1&1&0&0&1&0&0
\end{bmatrix}.
\end{align*}
This results in an \(\quantparam{8,1,3/3}\) code. Note that this procedure has also removed the $S_6^X$ and $S_2^Z$, $S_5^Z$, $S_6^Z$, and $S_7^Z$ stabilizers from the original code. Thus, the stabilizers left are the ones corresponding to $S_3^X$, $S_4^X$, $S_5^X$, $S_8^X$, $S_1^Z$, $S_3^Z$, and $S_4^Z$, but with the operators on qubits $1,2,6,8,9,10,11,12,13$ removed. 

\vspace{1mm}

\subsection{Encoding and switching between codes}
One of the advantages of using punctured codes instead of unrelated codes is that the sender can encode the qubit in the larger code and once they know which code they should use, they can easily modify the codeword to the punctured codeword by measuring specific qubits. Hence, the qubit is not required to go through an encoding circuit before it is ready for teleportation. The theory behind this, is that a puncturing is a projection, see \cite{Grassl21} Theorem 4.1 and \cite{Puncturing} Section III.D. Hence we can easily transform a codeword in the $\quantparam{17,1,5/5}$ code to a codeword in one of the shorter codes by projecting the qubits we want to remove to a fixed pure state. If we want to go from the $\quantparam{17,1,5/5}$ code to the $\quantparam{13,1,3/5}$ code, we puncture four qubits with respect to $(0|1)$. This is, as described in \cite{Puncturing}, equivalent to projecting the chosen qubits such that we have fixed these to be $\ket{0}$ before removing them. In the terminology of \cite{Grassl21} we are applying the projection $I\otimes \cdots \otimes I\otimes\ket{0}\bra{0}\otimes I\otimes \cdots \otimes I$, where $\ket{0}\bra{0}$ is on the index we want to remove. When going from the $\quantparam{13,1,3/5}$ code to the $\quantparam{8,1,3/3}$ code, we puncture with respect to $(1|0)$ for the given indices. This is equivalent to projecting these qubits to the state $\ket{+}$ before removing them.

If we cannot store the $17$ qubits in the memory, we still have an advantage over using unrelated codes. This is due to the relationship between the stabilizers of the large code and the punctured codes. The encoding circuit is only dependent on the stabilizers and since the stabilizers are related we can re-use parts of the encoding for the large code, if we want to encode in one of the punctured codes directly. Here we refer to the standard encoding scheme, see for instance \cite{Mondal}.

\subsection{Advantages of puncturing in the decoding procedure}
Due to the fact that we are keeping the same stabilizers but ignoring some qubits means that we can use parts of the same decoding circuit for the different codes. If, for instance, a codeword in the $\quantparam{13,1,3/5}$ code was received we can append the codeword with $\ket{0}$'s on the removed qubits and measure the eigenvalues of the stabilizers in the same way as we would have done in the $\quantparam{17,1,5/5}$ code. We can do this since the stabilizers we keep in order to obtain the $\quantparam{13,1,3/5}$ code will never have an $X$ stabilizer on the removed qubits. Hence, the kept stabilizers will stabilize the state after the $\ket{0}$'s are appended. We remark, that the eigenvalues for the removed stabilizers needs to be ignored when determining which error occurred. Similarly, if a codeword from the $\quantparam{8,1,3/3}$ code is received, we can append it with $\ket{+}$'s on the qubits punctured with respect to $(1|0)$ and $\ket{0}$'s on the qubits punctured with respect to $(0|1)$. Again, it can be sent through the same circuit for measuring the eigenvalues, but we only use the eigenvalues for the stabilizers $S_3^X$, $S_4^X$, $S_5^X$, $S_8^X$, $S_1^Z$, $S_3^Z$, and $S_4^Z$ to determine the error.

We remark, we do not have to measure all the eigenvalues of the stabilizers. Only the ones we will use to determine the errors. Hence, we can make it more efficient by removing parts of the decoding circuits. However, we want to emphasize that in any case, parts of the decoding circuit for the largest code can be re-used for the shorter codes when we apply puncturing due to the relation between the stabilizers of the codes.  

\begin{figure*}[t]
    \centering
    \begin{subfigure}{0.47\textwidth}
        \centering
        \includegraphics[width=\textwidth]{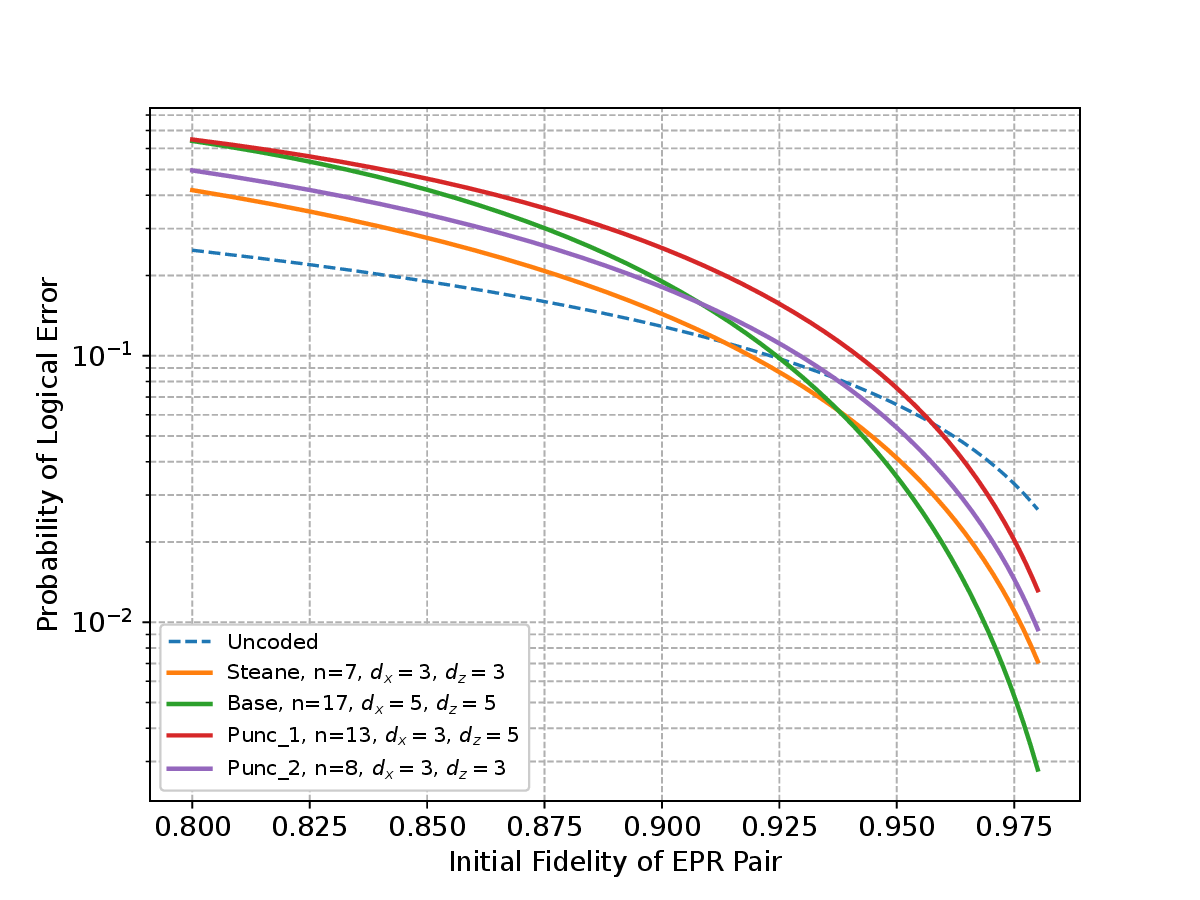}
        \caption{No purification ($r = 0$).}
        \label{fig:DEJMPS_0}
    \end{subfigure}
    \hfill
    \begin{subfigure}{0.47\textwidth}
        \centering
        \includegraphics[width=\textwidth]{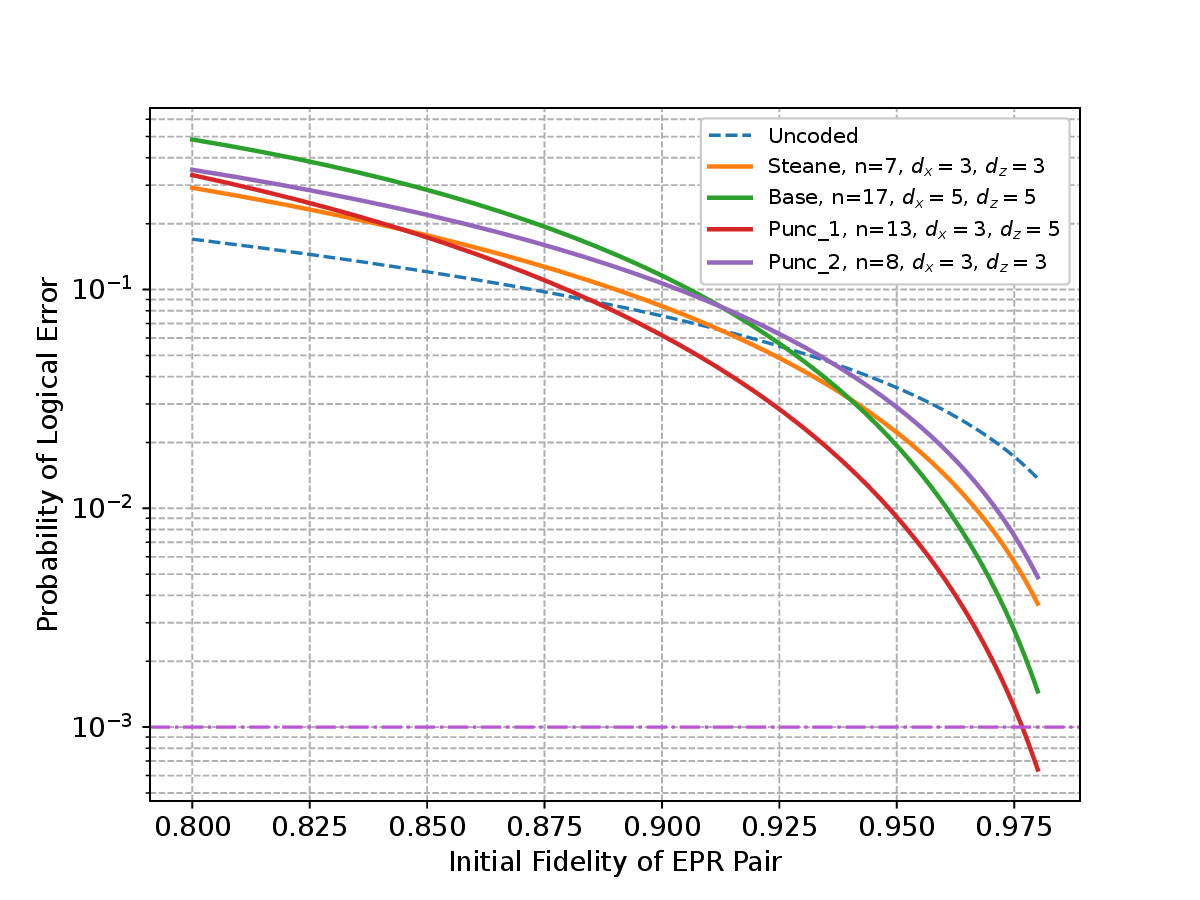}
        \caption{One purification round ($r = 1$).}
        \label{fig:DEJMPS_1}
    \end{subfigure}

    \begin{subfigure}{0.47\textwidth}
        \centering
        \includegraphics[width=\textwidth]{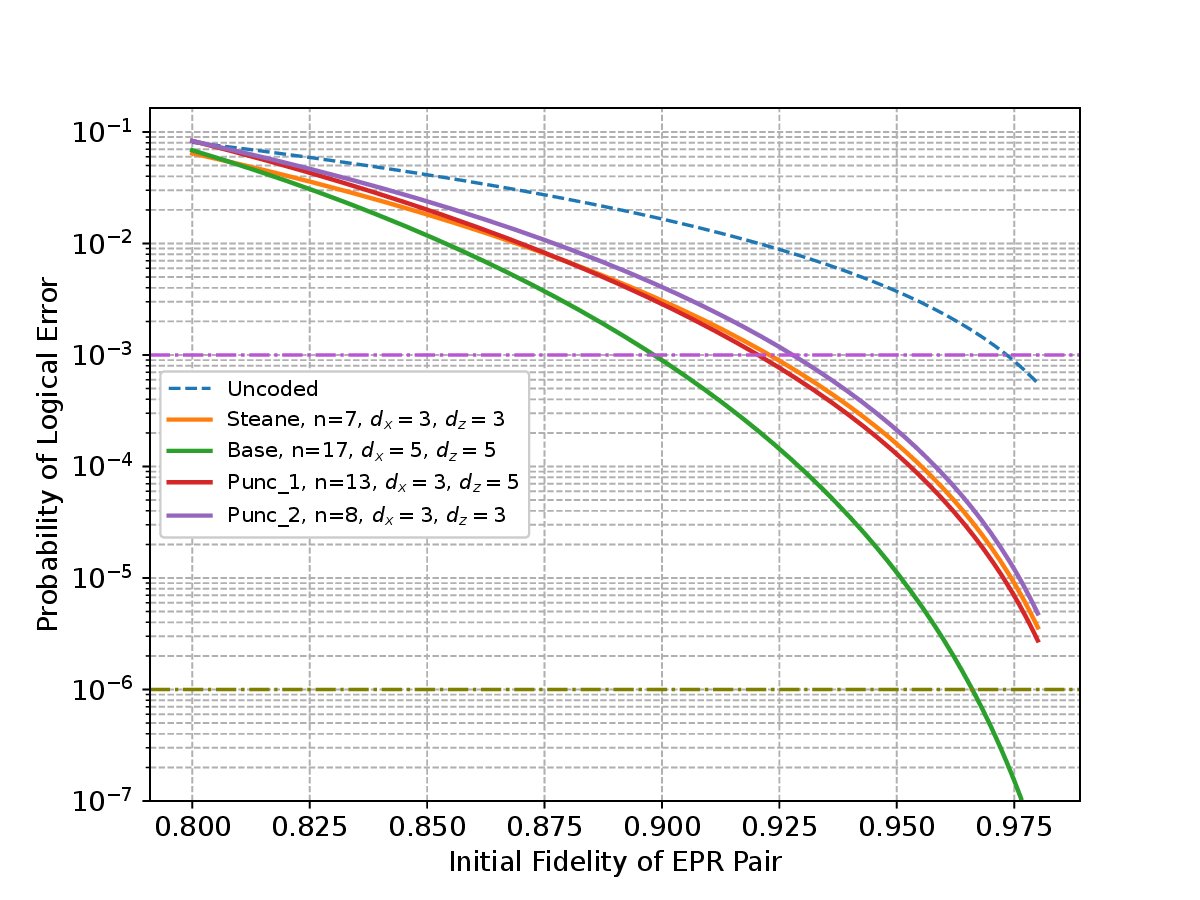}
        \caption{Two purification rounds ($r = 2$).}
        \label{fig:DEJMPS_2}
    \end{subfigure}
    \hfill
    \begin{subfigure}{0.47\textwidth}
        \centering
        \includegraphics[width=\textwidth]{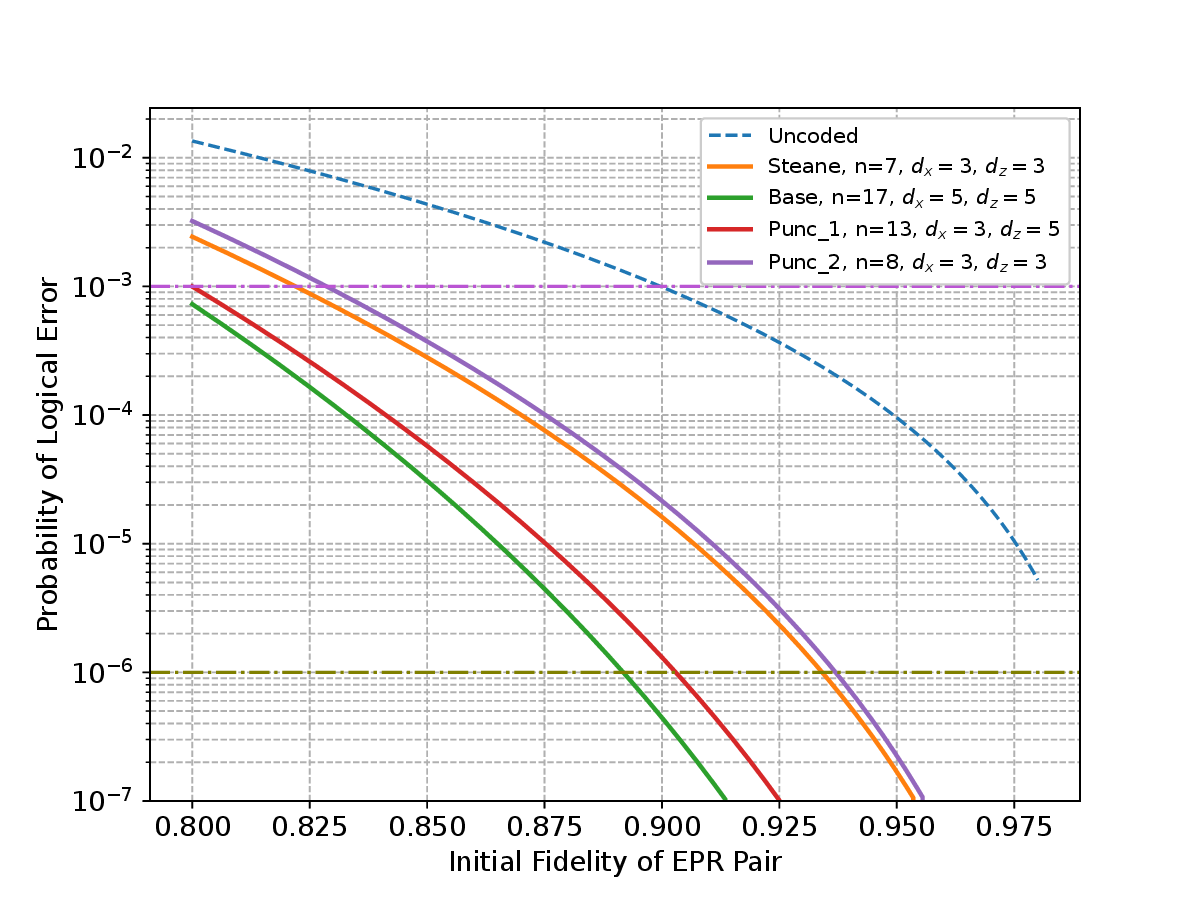}
        \caption{Three purification rounds ($r = 3$).}
        \label{fig:DEJMPS_3}
    \end{subfigure}
    \caption{Logical error probability versus initial EPR fidelity $F_0$ for DEJMPS purification with $r \in \{0,1,2,3\}$. Each panel compares the Steane code, the base CSS-coded scheme, and punctured variants, showing reduced dependence on high-fidelity entanglement and regime-dependent optimized puncturing. The horizontal lines at $10^{-3}$ and $10^{-6}$ indicate example reliability targets.}
    \label{fig:DEJMPS_full}
\end{figure*}
\section{Numerical Evaluations}
\label{sec:Numerical Evaluations}

This section presents a numerical evaluation of the base CSS code and its proposed punctured variants. Using the logical error probability of Section~\ref{modelling section}, each code is assessed under DEJMPS purification across varying initial EPR fidelities $F_0$ and purification rounds. The results demonstrate how coding with purification can meet target reliability with reduced initial fidelity requirements, and how puncturing further expands the feasible operating region by providing the best-performing variants across different $F_0$ values and purification rounds.

\subsection*{Codes Considered}
\begin{itemize}[leftmargin=*]
    \item \textbf{The \(\quantparam{17,1,5/5}\) CSS Code and Its Punctured Variants:}
          \begin{itemize}
              \item \(\quantparam{13,1,3/5}\): obtained by removing four qubits, preserving  phase flip distance while reducing bit flip distance,
              \item \(\quantparam{8,1,3/3}\): obtained by removing five additional qubits, yielding a shorter symmetric distance-three code.
          \end{itemize}
\item \textbf{Steane code \(\quantparam{7,1,3/3}\):}
\textcolor{black}{A symmetric CSS code correcting all single-qubit Pauli errors \cite{steane}.
As the shortest nondegenerate CSS code of distance~3, it serves as a standard benchmark for comparing CSS codes and their punctures.}

\end{itemize}

\subsection*{Numerical Results}
This section evaluates the logical error probabilities of different codes over a range of initial EPR fidelities $F_0$ and DEJMPS purification depths $r \in {0,1,2,3}$. The resulting reliability curves, computed using \eqref{eq:PL_css_corrected}, are shown in Fig.~\ref{fig:DEJMPS_full}. These scenarios illustrate how coding and purification interact and how puncturing provides additional flexibility across different operating regimes.

For the case where no purification is applied, Fig.~\ref{fig:DEJMPS_0} shows that for initial fidelities $F_0 \gtrsim 0.9$, coded teleportation yields a substantial reduction in logical error. The base $\quantparam{17,1,5/5}$ code achieves the lowest error, and even the shorter $\quantparam{8,1,3/3}$ puncture provides clear gains over the uncoded scheme. This indicates that coding alone can relax the fidelity requirements needed to achieve moderate reliability. In practice, certain target fidelities may be unattainable through purification due to timing constraints or decoherence during classical communication \cite{pur_with_latency_arxiv}. In such scenarios, the $\quantparam{17,1,5/5}$ code or the $\quantparam{8,1,3/3}$ puncture allows multiple lower fidelity pairs to achieve the desired reliability without relying on purification.

When one or two purification rounds are available, the benefit of puncturing becomes more pronounced. As shown in Fig.~\ref{fig:DEJMPS_1}, the $\quantparam{13,1,3/5}$ puncture matches or surpasses the base code at moderate and high fidelities. With a single purification round, it outperforms all other coded schemes for $F_0 \gtrsim 0.85$, including the longer $\quantparam{17,1,5/5}$ code. This reflects the asymmetric error structure introduced by DEJMPS purification, which suppresses some error types more strongly than others and thus favors codes, such as $\quantparam{13,1,3/5}$, that retain greater protection along the dominant $Z$ error branch.

The trends across Fig.~\ref{fig:DEJMPS_2} and Fig.~\ref{fig:DEJMPS_3} show that coding combined with shallow purification can achieve reliability levels that the uncoded scheme cannot match without substantially higher fidelity or deeper purification. For example, achieving a logical error below $10^{-3}$ requires the uncoded scheme to operate at fidelities above $0.975$ and to use two purification rounds, whereas the $\quantparam{13,1,3/5}$ puncture reaches this regime with only a single purification round. Moreover, for targets near $10^{-3}$ and $F_0 < 0.95$, the uncoded scheme requires three purification rounds, while the coded variants reach the same level with only two rounds; in this regime the shorter $\quantparam{8,1,3/3}$ puncture is often the most efficient choice due to its reduced blocklength.

With deeper purification ($r=3$), the shortest puncture becomes particularly competitive. Fig.~\ref{fig:DEJMPS_3} shows that the $\quantparam{8,1,3/3}$ code attains very low logical error at substantially lower fidelity than the uncoded scheme while using fewer qubits than the longer codes. For targets near $10^{-6}$, the $\quantparam{17,1,5/5}$ code reaches this level once $F_0 \gtrsim 0.9$, whereas the $\quantparam{8,1,3/3}$ puncture achieves the same reliability for $F_0 \gtrsim 0.94$. Notably, the $\quantparam{8,1,3/3}$ puncture, although slightly longer than the $\quantparam{7,1,3/3}$ Steane code, is obtained from the base code and therefore preserves compatibility with the same encoding/decoding circuit. This allows it to retain near-Steane performance in regions where the Steane code would be used while still providing the advantages of puncturing.

Overall, Fig.~\ref{fig:DEJMPS_full} demonstrates that no single code performs best across all conditions. Coding reduces the entanglement quality required to meet a reliability target, and puncturing provides a family of code variants that naturally adapt to the fidelity and characteristics of the channel. \textcolor{black}{In practice, for a given operating point, the logical error curves determine which punctured variants satisfy the reliability constraint, and the shortest such code is selected to minimize resource usage.} This adaptability enables reliable teleportation across a broader operating region without requiring hardware-level code switching.

\section{Conclusions \& Future Work}

Achieving reliable quantum teleportation in the presence of heterogeneous entanglement quality and reliability requirements is a key challenge in emerging systems. Toward this goal, this work investigated resource-adaptive encoded teleportation based on punctured CSS codes. By supplementing entanglement purification with punctured QEC codes, we derived a family of related codes by puncturing a single CSS code to obtain variants with different rates and $X/Z$ error-protection profiles, all sharing a common stabilizer structure.

Numerical results show that punctured codes improve teleportation reliability across a broader range of purification regimes, enabling target reliability without hardware-level code switching. Different punctured variants achieve the lowest logical error in different operating regimes, with asymmetric puncturing most effective when purification induces biased error channels. Selecting among these codes reduces logical error relative to fixed-rate encoded teleportation, lowering the initial EPR fidelity or number of purification rounds required to meet a given reliability target. Overall, puncturing enables adaptation to varying entanglement conditions and reliability requirements while reusing a single stabilizer structure.

Future work will explore how to balance purification and QEC under resource and timing constraints. This tradeoff depends on purification, code performance, and decoherence, motivating further study of joint purification and QEC.

\bibliographystyle{IEEEtran}
\bibliography{references.bib}

\end{document}